\documentclass[aps,prl,twocolumn,showpacs,superscriptaddress]{revtex4}
\usepackage{epsfig}
\usepackage{graphicx}
\usepackage{dcolumn}
\usepackage{amsmath}

\begin{document}


\title[Long Title]{Emergence of the N=16 shell gap in $^{21}$O}

\author{B. Fern\'andez-Dom\'inguez}
\affiliation{Oliver Lodge Laboratory, University of Liverpool, Liverpool L69 7ZE, UK}
\affiliation{GANIL, BP 55027, 14076 Caen Cedex 5, France}
\affiliation{LPC Caen, ENSICAEN, Universit\'e de Caen, CNRS/IN2P3, 14050 Caen, France}
\author{J. S. Thomas}
\affiliation{Department of Physics, University of Surrey, Guildford GU2 5XH, UK}
\author{W. N. Catford}
\affiliation{Department of Physics, University of Surrey, Guildford GU2 5XH, UK}
\author{F. Delaunay}
\affiliation{LPC Caen, ENSICAEN, Universit\'e de Caen, CNRS/IN2P3, 14050 Caen, France}
\author{S. M. Brown}
\affiliation{Department of Physics, University of Surrey, Guildford GU2 5XH, UK}
\author{N. A. Orr}
\affiliation{LPC Caen, ENSICAEN, Universit\'e de Caen, CNRS/IN2P3, 14050 Caen, France}
\author{M. Rejmund}
\affiliation{GANIL, BP 55027, 14076 Caen Cedex 5, France}
\author{N. L. Achouri}
\affiliation{LPC Caen, ENSICAEN, Universit\'e de Caen, CNRS/IN2P3, 14050 Caen, France}
\author{H. Al Falou}
\affiliation{LPC Caen, ENSICAEN, Universit\'e de Caen, CNRS/IN2P3, 14050 Caen, France}
\author{N. I. Ashwood}
\affiliation{School of Physics and Astronomy, University of Birmingham, Birmingham B15 2TT, UK}
\author{D. Beaumel}
\affiliation{Institut de Physique Nucl\'eaire, IN2P3/CNRS, 91406 Orsay Cedex, France}
\author{Y. Blumenfeld}
\affiliation{Institut de Physique Nucl\'eaire, IN2P3/CNRS, 91406 Orsay Cedex, France}
\author{B. A. Brown}
\affiliation{NSCL, Michigan State University East Lansing, MI 48824-1321, USA}
\author{R. Chapman}
\affiliation{School of Engineering and Science, University of the West of Scotland, Paisley PA1 2BE, UK}
\author{M. Chartier}
\affiliation{Oliver Lodge Laboratory, University of Liverpool, Liverpool L69 7ZE, UK}
\author{N. Curtis}
\affiliation{School of Physics and Astronomy, University of Birmingham, Birmingham B15 2TT, UK}
\author{C. Force}
\affiliation{GANIL, BP 55027, 14076 Caen Cedex 5, France}
\author{G. de France}
\affiliation{GANIL, BP 55027, 14076 Caen Cedex 5, France}
\author{S. Franchoo}
\affiliation{Institut de Physique Nucl\'eaire, IN2P3/CNRS, 91406 Orsay Cedex, France}
\author{J. Guillot}
\affiliation{Institut de Physique Nucl\'eaire, IN2P3/CNRS, 91406 Orsay Cedex, France}
\author{P. Haigh}
\affiliation{School of Physics and Astronomy, University of Birmingham, Birmingham B15 2TT, UK}
\author{F. Hammache}
\affiliation{Institut de Physique Nucl\'eaire, IN2P3/CNRS, 91406 Orsay Cedex, France}
\author{M. Labiche}
\affiliation{Nuclear Physics Group, STFC Daresbury Laboratory, Daresbury, Warrington WA4 4AD, UK}
\author{V. Lapoux}
\affiliation{IRFU, CEA-Saclay, 91191 Gif-sur-Yvette, France }
\author{R. C. Lemmon}
\affiliation{Nuclear Physics Group, STFC Daresbury Laboratory, Daresbury, Warrington WA4 4AD, UK}
\author{F. Mar\'echal}
\affiliation{Institut de Physique Nucl\'eaire, IN2P3/CNRS, 91406 Orsay Cedex, France}
\author{B. Martin}
\affiliation{IRFU, CEA-Saclay, 91191 Gif-sur-Yvette, France }
\author{A. Moro}
\affiliation{Departamento de F\'isica At\'omica, Molecular y Nuclear, Universidad de Sevilla, E-41080 Sevilla, Spain }
\author{X. Mougeot}
\affiliation{IRFU, CEA-Saclay, 91191 Gif-sur-Yvette, France }
\author{B. Mouginot}
\affiliation{Institut de Physique Nucl\'eaire, IN2P3/CNRS, 91406 Orsay Cedex, France}
\author{L. Nalpas}
\affiliation{IRFU, CEA-Saclay, 91191 Gif-sur-Yvette, France }
\author{A. Navin}
\affiliation{GANIL, BP 55027, 14076 Caen Cedex 5, France}
\author{N. Patterson}
\affiliation{Department of Physics, University of Surrey, Guildford GU2 5XH, UK}
\author{B. Pietras}
\affiliation{Oliver Lodge Laboratory, University of Liverpool, Liverpool L69 7ZE, UK}
\author{E. C. Pollacco}
\affiliation{IRFU, CEA-Saclay, 91191 Gif-sur-Yvette, France }
\author{A. Leprince}
\affiliation{LPC Caen, ENSICAEN, Universit\'e de Caen, CNRS/IN2P3, 14050 Caen, France}
\author{A. Ramus}
\affiliation{Institut de Physique Nucl\'eaire, IN2P3/CNRS, 91406 Orsay Cedex, France}
\author{J. A. Scarpaci}
\affiliation{Institut de Physique Nucl\'eaire, IN2P3/CNRS, 91406 Orsay Cedex, France}
\author{N. de S\'er\'eville}
\affiliation{Institut de Physique Nucl\'eaire, IN2P3/CNRS, 91406 Orsay Cedex, France}
\author{I. Stephan}
\affiliation{Institut de Physique Nucl\'eaire, IN2P3/CNRS, 91406 Orsay Cedex, France}
\author{O. Sorlin}
\affiliation{GANIL, BP 55027, 14076 Caen Cedex 5, France}
\author{G. L. Wilson}
\affiliation{Department of Physics, University of Surrey, Guildford GU2 5XH, UK}

\date{\today}

\begin{abstract}
The spectroscopy of $^{21}$O has been investigated using a radioactive $^{20}$O beam and the (d,p) reaction in inverse kinematics. The ground and first excited states have been determined to be $J^{\pi}$=5/2$^{+}$ and $J^{\pi}$=1/2$^{+}$ respectively. Two neutron unbound states were observed at excitation energies of 4.76 $\pm$ 0.10  and 6.16 $\pm$ 0.11. The spectroscopic factor deduced for the lower of these interpreted as a 3/2$^{+}$ level, reveals a rather pure $0d_{3/2}$  single-particle configuration.  The large energy difference between the 3/2$^{+}$ and 1/2$^+$ states is indicative of the emergence of the N=16 magic number.
For the higher lying resonance, which has a character consistent with a spin-parity assignment of 3/2$^{+}$ or 7/2$^{-}$, a 71\% branching ratio to the first 2$^{+}$ state in $^{20}$O has been observed. The results are compared with new shell model calculations.
\end{abstract}


\maketitle

The magic numbers that explain the structure of nuclei close to stability have their origin in the gaps created by the single-particle eigenstates of the mean-field. The structure of exotic nuclei is now known to often differ from that near stability, exhibiting an evolution of the shell closures or even a reordering of the levels \cite{sorlin-review}. The single-particle properties can be substantially modified in light neutron-rich nuclei by, in part, to the combined action of the central component and the tensor part of the effective nucleon-nucleon (NN) interaction \cite{smirnova10,otsuka10}. Furthermore, near the drip line, other effects become important including many-body correlations \cite{otsuka_3bf,hagen,hagen10}, and the influence of the scattering to the continuum in weakly bound and unbound states \cite{dobaczewski,tsukiyama,nmichel}.

The neutron-rich oxygen isotopes represent an intriguing means to study the interplay of such effects.  The last bound oxygen isotope is $^{24}$O, which reinforces N=16 as a shell gap, but the addition of a single proton moves the neutron-drip line for the fluorine isotopes to at least $^{31}$F (N=22) \cite{sakurai}. The N=16 gap is produced by an increase in the spacing between the $1s_{1/2}$  and the $0d_{3/2}$ neutron orbitals. A precise knowledge of how these orbitals evolve is crucial in order to predict the position of the neutron-drip line. Although fundamental in determining the binding energy of $^{26,28}$O, the energy of the $\nu0d_{3/2}$ single-particle orbital is still the object of disagreement between different shell model interactions. Several phenomenological interactions, such as the USDB \cite{usdab} and SDPF-M \cite{utsuno_sdpfm}, are able to successfully predict $^{24}$O as the last bound oxygen isotope. However, when derived from microscopic nucleon-nucleon (NN) forces, the position of the drip line is shifted to $^{28}$O \cite{otsuka_3bf}.  The inclusion of three-nucleon forces (3NFs) is known to influence the location of the drip line as pointed out first in ab-initio calculations for light nuclei by Hagen et al. \cite{hagen} and, more recently by Otsuka and collaborators \cite{otsuka_3bf}. The influence of continuum coupling, while being important in describing the low-lying states in $^{24}$O, does not appear to affect the position of the drip line \cite{tsukiyama}.

Previous work investigating the $\nu$0$d_{3/2}$ orbital employed transfer of a neutron onto $^{22}$O \cite{elekes}. The first 3/2$^{+}$ level in $^{23}$O was found at 4.00(2) MeV consistent with shell model calculations using the SDPF-M \cite{utsuno_sdpfm} and USDA \cite{usdab} interactions. In $^{25}$O, the location of the unbound $\nu$0$d_{3/2}$ ground state, observed in proton knockout from $^{26}$F \cite{hoffmanprl} could only be reproduced by the USD interaction \cite{brownsde} which in contrast does not predict correctly the neutron-drip line. In the present work we have employed the d($^{20}$O,p) reaction as a means to shed further light on the evolution of the $\nu0d_{3/2}$ strength.
In the naive shell model picture $^{20}$O has 4 neutrons in the $\nu$0$d5/2$ orbital. The stripping of a nucleon from the target is thus expected to populate the $\nu$0$d_{5/2}$, $\nu$1$s_{1/2}$ and $\nu$0$d_{3/2}$ orbitals. The size of the N=16 shell gap is related to the energy difference between the 3/2$^{+}$ and 1/2$^{+}$ states. Existing data on $^{21}$O \cite{wcatford21o,stanoiu} provide evidence for a low-lying 3/2$^{+}$ state, but this is expected to include significant core excitation rather than a strong $\nu0d_{3/2}$ single-particle structure. We report here on the measurement of the observation of a higher lying 3/2$^{+}$ level that carries significant $\nu0d_{3/2}$ strength.

\vspace{-0.4cm}
\begin{figure}[hb]
\begin{center}
\psfig{figure=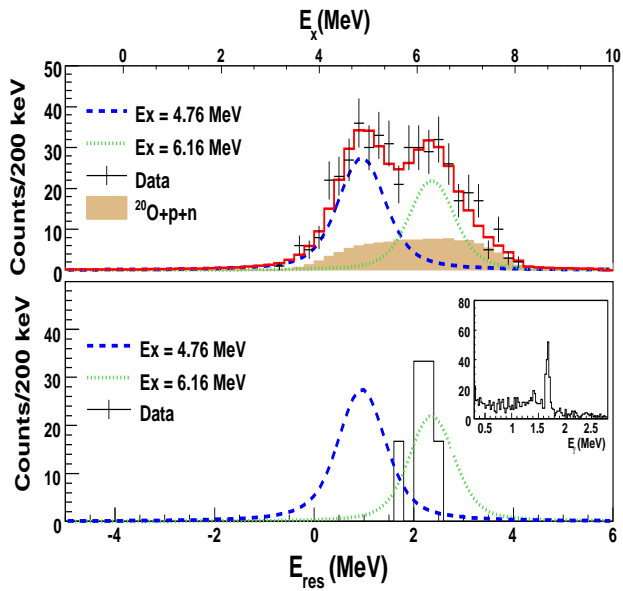,width=9.25cm,height=8.5cm,angle=0}
\end{center}
\caption{Upper: Barrel excitation energy spectrum of the unbound states in $^{21}$O from $^{20}$O+p coincidences. Protons were detected between [95-145] degrees. Bottom: Events corresponding to protons in coincidence with 1684 keV deexcitation $\gamma$ rays from $^{20}$O. The inset of the bottom figure shows the $\gamma$-ray spectrum gated on $^{20}$O and requiring an event in the TIARA Si array.}
\label{ex_unbd}
\end{figure}

The experiment was performed at the SPIRAL facility of GANIL in Caen using a 10.53 AMeV $^{20}$O beam with an average intensity of 10$^{4}$ pps.  The beam impinged on a 0.59 mg/cm$^{2}$ thick CD$_{2}$ target. A detailed description of the experimental set-up can be found in references \cite{labiche,wcatfordprl,wcatford05a}. The (d,p) reaction channel was selected using the identification of the beam-like residue using the VAMOS spectrometer and the kinematics of the proton in the position sensitive silicon array TIARA \cite{labiche} which surrounded (37$^\circ$-145$^\circ$) the target. The excitation energy in $^{21}$O and the centre-of-mass (cm) scattering angle was obtained from the measured energy and laboratory angle of the recoiling protons. The bound first-excited state of $^{21}$O was measured at 1.21 $\pm$ 0.04 MeV. As in our earlier work \cite{labiche,wcatfordprl,wcatford05a}, the target was also surrounded at 90$^\circ$ by 4 EXOGAM clover detectors in a close packed geometry (efficiency was 10\% at 1 MeV). The coincident gamma-energy spectrum provided a measurement of the excitation energy of $1.213\pm 0.007$ MeV. The values obtained both via the detection of the protons and the gamma ray compare well with previously established energy of 1.218 $\pm$ 0.004 \cite{stanoiu}. Any other bound states populated were over an order of magnitude weaker. 

The excitation energy spectrum deduced for states lying above the neutron emission threshold (Sn=3.806 $\pm$ 0.012 MeV) in $^{21}$O is shown in the upper part of Fig. \ref{ex_unbd}. The solid line represents the sum of two resonances (dashed and pointed lines) and a contribution from direct break-up ($^{20}$O+n+p).  The resonance line-shape employed was a Breit-Wigner  convolved with a Gaussian of fixed width (FWHM$_{exp}$) that represented the experimental resolution accounting for the beam characteristics (spot size and energy spread), the energy and angular straggling in the target, detector resolutions and kinematical effects. The three-body contribution, taken here to mean a background from the breakup of the deuteron ($^{20}$O+n+p), was obtained by uniformly sampling the phase-space of such a decay. The normalisation of this contribution was a free parameter in the fit. A Monte-Carlo simulation \cite{labiche} based on Geant4 was performed to determine the experimental resolution. The simulations were validated by comparison with the measured resolution of the state at 1.21 MeV, FWHM$_{exp}$=0.77 $\pm $0.17 MeV and  FWHM$_{sim}$=0.75 MeV. The simulations were then used to estimate the resolution at higher excitation energies. The adjustement to the data yielded  $E_{res_{1}}$ = 0.960 $\pm$ 0.10 MeV, $\Gamma_{1}$=0.560 $\pm$ 0.20, and $E_{res_{2}}$=2.36 $\pm$ 0.11 $\Gamma_{2}$=0.390 $\pm$ 0.44. The corresponding excitation energies, as listed in Table \ref{texpres}, of the unbound states are 4.76 $\pm$ 0.10 and 6.16 $\pm$ 0.11 MeV. The bottom part of Fig.\ref{ex_unbd} shows the excitation energy spectrum for the events for $\gamma$-particle-fragment triple coincidences, ($^{20}$O+p+$\gamma$). Taking into account the $\gamma$-ray efficiency, the second resonance was found to decay to the 1.684 MeV (2$^{+}$) first excited state in $^{20}$O with a branching ratio, $\Gamma(2^{+})/\Gamma_{tot}$, of $0.71\pm0.22$.

Shell model calculations presented in table \ref{tsmres} suggest that the states expected to be populated above the neutron threshold include the 3/2$^{+}$ and, possibly, negative parity states from the $fp$-shell. Any $\ell=1$ single-particle resonances at the measured excitation energies are expected to be very broad ($\Gamma_{sp}$(4.76)=3.4 MeV, $\Gamma_{sp}$(5.5)=100 MeV). Therefore, the measured widths of the observed resonances already excludes them from being  $\ell=1$ states with a reasonable spectroscopic factor.

\begin{figure}[h]
\centering
\psfig{figure=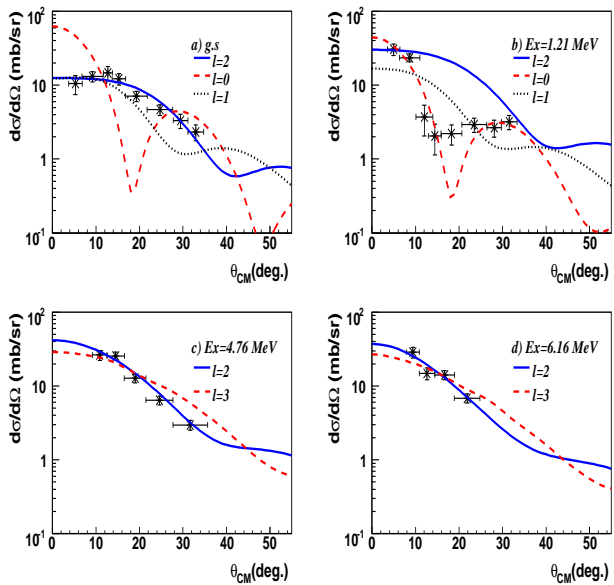,width=8.5cm,height=8cm,angle=0}
\caption{Differential cross sections for the bound a) and  b) and unbound c) and d) states in $^{21}$O compared to ADWA calculations. For the adopted $\ell$-transfers see the text and Table I. Only statistical uncertainties are included.}
\label{angdist}
\end{figure}

As in reference \cite{wcatfordprl}, elastic scattering was used to calibrate the beam current and number of deuterons in the target. The optical-model parameters obtained from d+$^{26}$Mg at 12 AMeV \cite{meurders74} were used and the final uncertainty in the normalisation was estimtated to be at most 2\%.

Figure \ref{angdist} displays the angular distributions for the protons leading to the bound and unbound states in $^{21}$O compared to reaction model calculations. The theoretical cross sections were computed within the Adiabatic Distorted Wave Approximation (ADWA) using an adiabatic potential for the d+$^{20}$O entrance channel \cite{johnson70}. The Chapel-Hill parameterisation (CH89) \cite{ch89} was used to describe the nucleon-nucleus potentials for both the entrance and outgoing channels.  All the calculations for the unbound states were performed using the Vincent and Fortune prescription \cite{vf} implemented in DWUCK4 \cite{dwuck}. The neutron single-particle form factors were obtained using a Woods-Saxon potential with the depth adjusted to reproduce the experimental binding energy of each level. Checks on the results for the unbound states were performed using bins in the continuum representing each resonance and implemented in the FRESCO code \cite{fresco}. Consistent results for the angular distributions were obtained with both approaches. The measured excitation energies, spectroscopic factors, and the assigned spins and parities are listed in Table I.

\begin{table}
\caption{Results from the present work for the states observed in $^{21}$O. BR=$\Gamma(2^{+})/\Gamma_{tot}$ for $^{20}$O+n decay.}

\begin{tabular}{ccc@{\hspace{0.4cm}}c@{\hspace{0.4cm}}c}
 \hline\hline
{} &{} &{} &{} &{} \\[-1.5ex]  

  $E_{x}$(keV) & ~$\ell$ ~& J$^{\pi}$ & BR & $C^{2}S$ \\ 

\hline 
\hline

  0 & 2      &$5/2^{+}$ & - & 0.34 $\pm$ 0.03\\
  1213 $\pm$ 7 & 0   &$1/2^{+}$ & - & 0.77 $\pm$ 0.09\\ 
  4760 $\pm$ 100 & 2   &$3/2^{+}$ & - & 0.58 $\pm$ 0.06\\
  6160 $\pm$ 110 & 2 & $3/2^{+}$ & 0.71$\pm$0.22 & 0.30 $\pm$ 0.05\\ 
                 & 3 & $7/2^{-}$ & 0.71$\pm$0.22 & 0.20 $\pm$ 0.02\\
\hline 
\end{tabular}
\label{texpres}
\end{table}

The angular distribution for the transfer to the ground state is well reproduced by an $\ell=2$ transfer, in agreement with a 5/2$^{+}$ assignment made in earlier studies \cite{wcatford21o,stanoiu}. The differential cross sections for the 1.21 MeV state are only consistent with an $\ell=0$ transfer, which permits a definite 1/2$^{+}$ assignment to this state. The spectroscopic factors were obtained by minimising the $\chi^{2}$ for all the data points. The spectroscopic factors of the ground and first excited states were determined to be 0.34 $\pm$ 0.03 and 0.77 $\pm$ 0.09, respectively, in agreement with the predictions using the USDA, SDPF-M and WBP interactions (Table \ref{tsmres}).  The 5/2$^{+}$ and 1/2$^{+}$ bound states carry most of the available strength of the $0d_{5/2}$ and $1s_{1/2}$ orbitals, respectively, based on the vacancies of the projectile. The occupancy of the $1s_{1/2}$ orbital is smaller than unity as would be expected in a naive shell model picture. This is consistent with this orbital being partially occupied in the projectile wave function as shown by a complementary study of the $^{20}$O(d,t)$^{19}$O reaction \cite{aramus_proc}. 

The angular distribution for the resonant state at E$_{x}$=4.76 MeV is best reproduced by an $\ell=2$ transition, leading to a spin-parity assignment of $J^{\pi}$=(3/2$^{+}$, 5/2$^{+}$). Comparisons with shell model calculations using different interactions predict no significant spectroscopic factor strength  for a 5/2$^{+}$ state in this energy region (Table \ref{tsmres}). States with negative parity  $fp$-configurations are predicted to lie 1 MeV higher in excitation energy than the strong, $\nu0d_{3/2}$, 3/2$^{+}$ state by the WBP and SDPF-M interactions. Therefore, it is concluded that the state at 4.76 MeV is very likely 3/2$^{+}$. The corresponding experimental spectroscopic factor of 0.58 $\pm$ 0.06 for $\ell=2$ is in very good agreement with the USDA value of 0.68, while, as listed in Table \ref{tsmres}, the WBP interaction predicts this state to carry a somewhat larger fraction of the $0d_{3/2}$ single-particle strength.  The SDPF-M interaction predicts the corresponding state to have a very low spectroscopic factor and with the strength concentrated in a higher lying state.

\begin{table*}
\caption{Shell Model calculations for $^{21}$O using the USDA/SDPF-M/WBP interactions for the $sd-pf$ shell states.}
\begin{tabular}{cc@{\hspace*{0.4cm}}c@{\hspace*{0.8cm}}cc@{\hspace*{0.8cm}}cc@{\hspace*{0.8cm}}cc}
  \hline\hline
{} &{}  &{} &{} &{}  &{} &{} &{} &{}\\[-1.5ex]
  J$^{\pi}$ & $(nlj)$ & $i$ & \multicolumn{2}{c}{USDA \cite{usdab}} & \multicolumn{2}{c}{SDPF-M \cite{utsuno_sdpfm}} &\multicolumn{2}{c}{WBP \cite{wbp}}  \\
    & & & $E_{x}$(keV) & $C^{2}S$ & $E_{x}$(keV) & $C^{2}S$ & $E_{x}$(keV) & $C^{2}S$  \\[0.5ex]
\hline \hline

  1/2$^{+}$ & 1$s_{1/2}$& 1 &1277 & 0.83 &  1451 & 0.82 &  1331 & 0.81 \\
            &           & 2 &6981 & 0.00 &  6695 & 0.00    &  5988 &
            0.00
            \\ \hline

  3/2$^{+}$ & 0$d_{3/2}$& 1& 2186 & 0.00  & 1926 & 0.00   & 2189 & 0.00 \\
            &           & 2& 5522 & 0.68  & 5446 & 0.08    & 4836 & 0.82 \\
            &           & 3& 5671 & 0.13  & 6008 & 0.77    & 5796 & 0.03 \\
            &           & 4& 8195 & 0.11  & 8666 & 0.09    & 7897 & 0.06 \\
\hline

  5/2$^{+}$ & 0$d_{5/2}$& 1& 0 & 0.34 &  0 & 0.33 & 0.0&  0.34 \\
            &           & 2& 3129 & 0.04 &  3034 & 0.04 &  3149 & 0.04 \\
            &           & 3& 4856 & 0.00 & 4796 & 0.00 &  4965 & 0.00 \\
            &           & 4& 7537 & 0.01 &  7296 & 0.00 &  6547 & 0.01 \\

\hline
  3/2$^{-}$ & 1$p_{3/2}$& 1& - & - &  7003 & 0.48 &  6223&0.36 \\
            &           & 2& - & - &  8427 & 0.15 &  6905& 0.10 \\
            &           & 3& - & - &  8760 & 0.00    &  7481& 0.02  \\
\hline

  7/2$^{-}$ & 1$f_{7/2}$& 1& - & - &  6812 & 0.70&  5942& 0.12  \\
            &           & 2& - & - &  8698 & 0.11&  6469& 0.05  \\

\hline
\end{tabular}
\label{tsmres}
\end{table*}

For the second unbound state at 6.16 MeV, the angular distributions for $\ell=2$ and $3$ transfer agree with the data equally. The corresponding spectroscopic factors are 0.30$\pm$0.05 and 0.20$\pm$0.02. For $\ell=1$, no resonance can be generated this far above the threshold, as such, a very small spectroscopic factor would, therefore, be needed for the width of an $\ell=1$ state to be consistent with the observed width of 0.39 MeV. As the decay of the 6.16 MeV state to the ground state represents only 0.30 $\pm$ 0.22 of the decay, an $\ell=1$, (3/2$^{-}$) assignment is very unlikely. From Table \ref{tsmres}, the states expected to be strongly populated in the (d,p) reaction can be identified. Leaving aside the 3/2$_{2}$$^{+}$ (3/2$_{3}$$^{+}$ state in SDPF-M) identified with the 4.76 MeV state, the remaining candidates are the other 3/2$^{+}$ states together with the 7/2$^{-}$ and 3/2$^{-}$ states. For all of these, the energy gap is 1.5 MeV  relative to the 3/2$_{2}$$^{+}$. On the other hand the only states in Table \ref{tsmres} favouring decay to the $^{20}$O(2$^{+}$) turn out to be the 3/2$_{4}$$^{+}$, the 7/2$_{2}$$^{-}$ and the 3/2$_{3}$$^{-}$ states. For the 3/2$_{4}$$^{+}$, for example, the three interactions predict values of 0.69, 0.74 and 0.76 for $\Gamma(2^{+})/\Gamma_{tot}$, but are associated with a spectroscopic factor of only 0.1 (Table \ref{tsmres}). It seems, therefore, that the shell model does not mix the various configurations in the 3/2$^{+}$ states correctly. Assuming the state at 6.16 MeV to be $J^{\pi}=3/2^{+}$, then the  $0d_{3/2}$ strength would be split between the 4.76 and 6.16 MeV states in a manner that is not predicted by the calculations. The other possibility is a $7/2^{-}$ assigment, with a wavefunction carrying a small part of the $\nu0f_{7/2}$ strength and dominated by core excitations coupled to a $\nu1p_{3/2}$ neutron.

As noted earlier, an estimate of the limit of the size of the N=16 shell gap in $^{21}$O can be deduced from the difference in excitation energies of the 1/2$^{+}$ (Ex=1.21 MeV) and the 3/2$^{+}$ (Ex=4.76 MeV) states. The value here of 3.5 MeV is close to that of 4 MeV in  $^{23}$O \cite{elekes}.  The invariant mass spectra of the unbound states in $^{23}$O populated in the d($^{22}$O,p) reaction also shows a second resonance at roughly 1.3 MeV distance to the first one compare to ours at 1.4 MeV. A 3/2$^{-}$ assignment was proposed for this state \cite{elekes}. However, the width of the resonance is too narrow to be in agreement with the spectroscopic factor of 1 deduced from the magnitude of the cross section. This inconsistency might arise from the weakly bound approximation used for the final state in the reaction calculations.

Summarising, we have presented results for neutron transfer populating the bound and  unbound states in $^{21}$O. The first excited state was assigned spin parity 1/2$^{+}$ and the $\ell=2$ transfer to the ground state supports a 5/2$^{+}$ assignment. The spectroscopic factors derived for both states are in good agreement with the shell-model predictions. Two neutron unbound states have been observed at 4.76 and 6.16 MeV. The 4.76 MeV state is identified as 3/2$^{+}$ and, as such, carries ~ 58\% of the $\nu0d_{3/2}$ strength. The state at 6.16 MeV corresponds either to a large fraction of the missing $\nu0d_{3/2}$ strength or to an intruder $f$-state from the $pf$-shell. The shell model is unable to provide an adequate explanation for this level. This poses a challenge for the shell model and the interactions used to describe the evolution of nuclear structure in this region. Ideally the issues discussed here should be investigated using less phenomenological models that include explicitely the physics of interest, such as the continuum. The triple-coincidence technique has been shown to be a very useful tool to probe the structure of unbound states with sizeable core excitations components. This suggests the equipment being developed at emerging ISOL facilities for studying transfer reactions should incorporate highly efficient $\gamma$-ray arrays, even when studying unbound states. 
 
The authors acknowledge the excellent support provided by the technical staff of LPC and GANIL and wish to thank Dr. N. Timofeyuk and Profs. J. Camacho, J. A. Tostevin and Y. Utsuno for fruitful discussions. This work was supported partially by the European Union Sixth Framework through the EURISOL Design-Study, Contract no. 515768 RIDS.

\end{document}